\documentclass[12pt,english,a4paper]{article}
\pdfoutput=1

\usepackage{mathtools,amssymb,tabu,babel,cite,slashed,heptit}
\usepackage[width=15.5cm,height=23.5cm,centering]{geometry}
\usepackage{setspace}
\usepackage[hidelinks,pagebackref]{hyperref}
\usepackage{cleveref}

\numberwithin{equation}{section}

\newcommand\be{\begin{equation}\begin{aligned}}
\newcommand\ee{\end{aligned}\end{equation}}

\newcommand\mathown{\mathbb}
\newcommand\CP{\mathown {CP}}
\newcommand\RP{\mathown {RP}}
\newcommand\IR{\mathown R}
\newcommand\IC{\mathown C}
\newcommand\IZ{\mathown Z}
\newcommand\IX{\mathown X}
\newcommand\NN{\mathcal N}

\let\Im\relax \DeclareMathOperator{\Im}{Im}
\let\Re\relax \DeclareMathOperator{\Re}{Re}
\DeclareMathOperator\pco{PCO}
\DeclareMathOperator\ch{Ch}
\renewcommand\tilde\widetilde
\renewcommand\hat\widehat
\renewcommand\bar\overline

\title{Real topological string amplitudes}

\author[a]{K.S.~Narain}
\author[b,c]{N.~Piazzalunga}
\author[c]{A.~Tanzini}

\affil[a]{The Abdus Salam International Centre for Theoretical Physics (ICTP)\affilcr
strada Costiera 11, 34151 Trieste, Italy}
\affil[b]{Simons Center for Geometry and Physics, State University of New York\affilcr
Stony Brook NY 11794-3636, USA}
\affil[c]{International School for Advanced Studies (SISSA) and INFN, Sez.~di Trieste\affilcr
via Bonomea 265, 34136 Trieste, Italy}

\abstract{
We discuss the physical superstring correlation functions in type I theory
(or equivalently type II with orientifold)
that compute real topological string amplitudes.
We consider the correlator corresponding to \emph{holomorphic} derivative
of the real topological amplitude $\mathcal G_\chi$,
at fixed worldsheet Euler characteristic $\chi$.
This corresponds in the low-energy effective action to $\NN=2$ Weyl multiplet,
appropriately reduced to the orientifold invariant part, and raised to the power $g'=-\chi+1$.
We show that the physical string correlator
gives precisely the holomorphic derivative of topological amplitude.
Finally, we apply this method to the standard closed oriented case as well,
and prove a similar statement for the topological amplitude $\mathcal F_g$.
}

\begin{document}

\maketitle

\tableofcontents

\section{Introduction}

Topological string amplitudes have a wide range of applications
both for the study of low-energy effective actions arising in superstring compactifications
and for their clear connection with mathematics.
For these reasons, they represent a very effective playground
where to test and improve our ability to calculate superstring amplitudes.
The system we consider in this paper, dubbed real topological string\cite
{Walcher:2007qp,Krefl:2009mw,Krefl:2009md,Krefl:2008sj,Piazzalunga:2014waa},
may be approached either from type IIA or type I perspective.
From type IIA viewpoint, we start from a Calabi-Yau threefold $\IX$
that admits an anti-holomorphic involution $\sigma$
with non-empty fixed point set $L$, which is therefore Lagrangian and supports an orientifold plane.
If we are interested in the topological subsector,
the spacetime directions spanned by the orientifold are not constrained a priori,
but can be guessed in the following way:
if we take the O-plane to span $L$ and two spacetime directions,
namely consider also a spacetime involution $x^2, x^3 \to - x^2, -x^3$,
then the O4-plane we get is such that we can cancel its charge
by wrapping precisely one D4-brane on the same locus, as seen in the covering.
Such local tadpole cancellation condition
on one hand guarantees integrality of the BPS expansion one gets
from combining the various unoriented and open topological amplitudes,
on the other hand it is the topological analogue of physical tadpole cancellation.
Moreover, it is related to decoupling
of vector- and hyper-multiplet in spacetime \cite{Bonelli:2009aw}.
We take as definition of the real topological string
the counting of holomorphic maps from Riemann surfaces to $\IX$
that are equivariant w.r.t.\ $\sigma$ and worldsheet parity $\Omega$,
subject to local tadpole cancellation.\footnote
{In some concrete examples,
this amounts to a constraint relating the Euler characteristic of the surface to the degree of the map.}

\paragraph{}

For simplicity we shall consider in our calculations an orbifold limit of a Calabi-Yau space,
with orbifold of the form $(T^2)^3/G$,
$G$ being the orbifold group,
for example $G=\IZ_2 \times \IZ_2$.
However, our results can be generalized to an arbitrary CY threefold by using CFT arguments\cite{Antoniadis:1993ze}.
Let $(Z^3,Z^4,Z^5)$ denote the complex coordinates on the three $T^2$s.
We take the D4-O4 system to be along $x^0,x^1$ directions and wrapping a Lagrangian subspace
along $r(i) = \Re Z^i$ for $i=3,4,5$.
Defining complex coordinates for spacetime as $Z^1=x^0+i x^2$ and $Z^2=x^1+i x^3$
and the left and right moving fermionic partners of $Z^i$ as $\psi^i$ and $\tilde{\psi^i}$
we conclude that orientifolding (which takes type II to type I-like theory)
is defined by the action of world sheet parity operator $\Omega$
(exchanging $\psi$ and $\tilde{\psi}$)
combined with $\IZ_2$ involution
\be
\label{involution}
\sigma: (Z^i,\psi^i,\tilde{\psi^i}) \mapsto (\bar{Z^i},\bar{\psi^i},\bar{\tilde{\psi^i}}),
\quad i=1,\ldots,5
\ee

This particular model (in its simplest version) can be obtained starting
from the standard Type I theory
(i.e.\ IIB quotiented by $\Omega$ with the associated 32 D9-branes and O9-plane)
compactified on $\IX$.\footnote
{In more complicated constructions one can also have D5-branes and O5-planes
wrapping various 2-cycles of $\IX$.
However this will not change the closed string sector,
which is what we are concerned with in trying to identify the self-dual background.}
One can get our model by doing 5 T-dualities,
3 of them inside $\IX$ so that one gets the mirror CY
and the remaining 2 T-dualities on $x^2$ and $x^3$ directions.
This results, by the standard rules of T-duality, in the presence of world-sheet parity projection
$ \Omega  \to  \Omega.\sigma$
and it converts original O9 and D9-branes into O4 and D4-branes
along the $\sigma$-fixed point set,
i.e.\ for the orbifold realization of $\IX$ along $(x^0, x^1, \Re Z^3, \Re Z^4, \Re Z^5)$.

\subsection{Useful facts}

Every surface $\Sigma_{g,h,c}$,
where $(g,h,c)$ denote respectively the number of handles, boundaries and crosscaps,\footnote
{A crosscap is topologically $\RP^2$.}
can be realized as a quotient $\Sigma_{g'}/\Omega$,
where $\Sigma_{g'}$ is Riemann surface of genus $g'$,
and $\Omega$ an orientation reversing involution.
The Euler characteristic of $\Sigma_{g,h,c}$ is $\chi =1-g'= 2- 2g -h -c$.

Let $h=h(\Sigma_{g'},\Omega) \geq 0$ be the number of components
of the fixed point set $\Sigma_{g'}^\Omega$ of $\Omega$ in $\Sigma_{g'}$
(i.e.\ the number of boundaries of $\Sigma_{g'}/\Omega$),
and the index of orientability
$k=k(\Sigma_{g'},\Omega)=\left( 2 - \# \text{ components of }\Sigma_{g'} \setminus \Sigma_{g'}^\Omega \right)$.
Then the topological invariants $h$ and $k$ together with $g'$
determine the topological type of $\Sigma_{g'}/\Omega$ uniquely.
For fixed genus $g'$, these invariants satisfy\cite{weichold}
\begin{itemize}
\item $k=0$ or $k=1$ (corresponding to oriented surfaces, or otherwise)
\item if $k=0$, then $0 < h \leq g'+1$ and $h \equiv g'+1 \mod{2}$
\item if $k=1$ then $h \leq g'$.
\end{itemize}
Moreover, the total number of topologically distinct surfaces at fixed $g'$ is given by
$\lfloor \frac{-3\chi + 8}2 \rfloor$,
where for even $\chi$ we also included in the count the case where $\Sigma_{g,h,c}$ itself is a closed oriented Riemann surface,
which is not of the form $\Sigma_{g'}/\Omega$ for $\Sigma_{g'}$ connected.

It is useful to separate the worldsheets $\Sigma_{g,h,c}$ into three classes,
corresponding to having 0, 1 or 2 crosscaps.\footnote
{Two crosscaps are equivalent to a Klein handle,
namely two holes glued together with an orientation reversal,
which in the presence of a third crosscap can be turned into an ordinary handle.}
This leads to a split of the topological string amplitudes into classes, namely
oriented surfaces with $h \geq 0$ boundaries
with amplitude $\mathcal F_{g,h}$,
non-orientable surfaces with an odd number of crosscaps
with amplitude $\mathcal R_{g,h}$,
and non-orientable surfaces with an even number of crosscaps
with amplitude $\mathcal K_{g,h}$.
We will be interested in their sum at fixed $\chi$, namely
\be
\label{total-g-chi}
\mathcal G_\chi = \frac12 \left(
\mathcal F_{g_\chi} + \sum_{2-2g-h=\chi} \mathcal F_{g,h} + \sum_{2-2g-h=\chi} \mathcal K_{g,h} +
\sum_{1-2g-h=\chi} \mathcal R_{g,h}
\right)
\ee
where we singled out the $h=0=c$ case, denoted simply by $\mathcal F_{g_\chi}$ with $g_\chi=1-\chi/2$.

\paragraph{Remark on relative homology}
In real topological string and Gromov-Witten theory,
one considers maps that are equivariant with respect to worldsheet involution $\Omega$ and target involution $\sigma$.
Such maps are classified by the second homology group:
although for purely open subsector it makes sense to consider relative homology $H_2(\IX,L;\IZ)$,
namely to reduce modulo 2-cycles supported on the Lagrangian $L$,
when dealing with the real theory it is not sufficient mathematically to specify the relative homology,
rather the full second homology is needed,
as one is formally studying maps from the symmetric surface to $\IX$.
In some simple cases, like the real quintic or local $\CP^2$, the two objects are isomorphic.

\paragraph{Remark on K\"ahler moduli}
Type IIA orientifold action has a well-defined action on the
moduli space $\mathcal M^k \times \mathcal M^q$\cite{Grimm:2004ua}:
the relevant subspace $\tilde {\mathcal M}^k$ of $\mathcal M^k$ is a special K\"ahler submanifold
of dimension $h^{1,1}_-$,
namely if we denote the K\"ahler form $J$ the action $\sigma^* J = -J$ induces a decomposition $h^{1,1}=h^{1,1}_+ +h^{1,1}_-$.
The simplest examples have $h^{1,1}=h^{1,1}_-=1$.

\subsection{What we compute}
\label{what-compute}

The mathematical foundations of real GW theory have been recently put on a solid ground\cite
{Georgieva:2015tax,Georgieva:2015sda,Georgieva:2015dha},
in particular we know that in constructing a nice moduli space of maps
one should not restrict to a single topological type of surface,
but rather consider them at once, morally as in \cref{total-g-chi},
so that orientation issues along real codimension one boundaries in moduli space are taken care of,
and one can construct a well-defined enumerative problem.

We are interested in the target space interpretation of topological strings.
Here, in the context of standard $\mathcal F_g$ in the oriented type II theory,
a study on the Heterotic dual\cite{Antoniadis:1995zn} gave the Schwinger formula
describing the singularity structure that explicitly proved the $c=1$ conjecture
and was generalized to all BPS states by Gopakumar and Vafa\cite{Gopakumar:1998ii,Gopakumar:1998jq}.
Recently their work has been revisited and clarified in the paper\cite{Dedushenko:2014nya},
which we point to for references and details.
A related aspect of target space physics connection to topological string amplitudes
is the well-known fact\cite{Antoniadis:1993ze} that
topological strings compute certain corrections to superstring amplitudes.
In order to understand how this result can be extended to the real setup, which is our goal,
it is useful to briefly recall some facts about the standard closed oriented case.

\paragraph{}

The physical string amplitude that computes $\mathcal F_g$ in type II theories
was identified\cite{Antoniadis:1993ze} with the F-term coupling
$\mathcal F_g (\mathcal W^2)^g$ in the low energy effective action,
where $\mathcal W$ is the chiral Weyl superfield.
Expanding in component fields one gets a term
proportional to $\mathcal F_g R^2 (T^2)^{g-1}$,
with $R$ and $T$ being respectively the anti-self-dual Riemann tensor and graviphoton field strength.
In the resulting amplitude there were two terms:
one where the fermionic bilinear terms of the graviton vertex operator contribute,
and the other where the bosonic part of the graviton vertex contributes.
For the fermionic term, by choosing a particular gauge for the positions of picture changing operators (PCO),
it was possible to carry out the spin structure sum and the result was shown to produce the topological string amplitudes.
The bosonic term was shown to vanish in a different gauge choice.
While the final answer must be true,
the choice of a different gauge for the two terms is unsatisfactory,
as in general, the position independence of PCO requires that all the operators are BRST invariant
while the fermionic and bosonic part of the graviton vertex operator are not separately BRST invariant.
It turns out however, that for the amplitude considered
in \cite{Antoniadis:1993ze}, the position independence of PCO holds separately for the bosonic and fermionic parts of the graviton vertices.\footnote
{We thank Ashoke Sen\cite{Ashoke} for pointing this out.}
This can be seen by writing one of the PCO as $\oint j_{BRST} \xi$ and deforming the contour.
Since only $\gamma G^-$ part of $j_{BRST}$ contributes to the amplitude
due to balancing of $U(1)$ charge, it annihilates separately the bosonic and fermionic parts
of the graviton vertex and therefore one can choose different gauges for the two parts.
In \cref{app93}, we give an alternative derivation of the result\cite{Antoniadis:1993ze}
by computing the term $(D_a \mathcal F_g) (F_a.R.T) (T^2)^{g-1}$,
where the subscript $a$ denotes a vector multiplet and $F_a$ its anti-self-dual field strength.
This is obtained from the superfield $\mathcal F_g (\mathcal W^2)^g$
by extracting two $\theta$s from $\mathcal F_g$.
The resulting amplitude involves only the fermionic bilinear term of the graviton vertex,
and with a suitable gauge choice one can perform the spin structure sum
and show that the result gives the holomorphic derivative of the topological string amplitude
along the $a$ direction.
This has the advantage that one can deal with both $g=1$ and $g>1$ uniformly, as for $g=1$,
direct evaluation of $\mathcal F_1$, namely $R^2$, vanishes on-shell
while the evaluation of $D_a \mathcal F_1$, i.e.\ $F_a. R. T$ is on-shell non-zero.

\paragraph{}

In the real case, we expect that the topological string computes the coupling of the form $RT^{g'-1}$
(in the covering picture $1-g'=\chi$).
In the following, to avoid the problem of having two different terms contributing to the amplitude
that require different gauge fixings,
we will again consider an amplitude involving a certain number of anti-self-dual graviphotons
and one matter anti-self-dual field strength,
namely we will compute the term involving one holomorphic covariant derivative with respect to a K\"aher modulus
of the real topological amplitude
that appears in the expansion of the orientifold invariant part of Weyl supergravity tensor.

\paragraph{Organization}

\Cref{real-top-amp-sec} deals with the superstring correlation functions that compute real topological string amplitudes.
In order to make the calculations explicit and transparent we consider
orbifold examples in \cref{projection}.
In \cref{sugra}, we discuss, from the supergravity perspective,
the possible low energy effective action term that the real topological string computes.
In \cref{inv-vert}, we describe the relevant orientifold invariant vertices
and the amplitudes are computed in full detail in \cref{Computation}.
The general case, namely smooth Calabi-Yau case, is briefly discussed in \cref{General CY}
and the special case of $g'=1$ in \cref{g'=1}.
\Cref{conclusion} lists some open questions and offers our conclusions.
\Cref{app93} presents the same method applied to the standard closed oriented case,
while \cref{theta} lists some useful facts about theta functions.

\section{Type II with orientifold}
\label{real-top-amp-sec}

\subsection{Projection}
\label{projection}

To get the Ramond-Ramond operators that survive orientifold projection, the easiest way is to
start from original type I compactified on $\IX$, which for concreteness
we take to be $(T^2)^3/(\IZ_2)^2$ (Gimon-Polchinsky model\cite{Gimon:1996rq,Pradisi:1988xd}).
In type I the only R-R field is the 3-form field strength
and its vertex operator in $(-1/2, -1/2)$ ghost picture is
\be
T_{MNP} = \tilde S^t.C.\Gamma_{MNP} S
\ee
where $S$ and $\tilde S$ are left and right moving 10d spin fields,
$C$ is the charge conjugation matrix and $M,N,P$ are 10d indices.
Orbifold projection $(T^2)^3/(\IZ_2)^2$ implies that
\be
(MNP) = (\mu \nu \rho), (\mu i j), (i j k)
\ee
where indices $i,j,k$ in $(i j k)$ are coming one index from each $T^2$
and in $(\mu i j)$ $i$ and $j$ come from the same $T^2$
(this ensures orbifold group invariance).
We can do the 5 T-dualities on these R-R operators:
they are given by the corresponding 5 parity actions
(i.e.\ action of $\sigma$ defined in \cref{involution})
on $S$
(but not on right moving $\tilde S$ --- this is because T-duality can be thought of
as parity transformation only on left movers):
\be
S \to \Gamma_{23 i(a) i(b) i(c)} S
\ee
where $i(a), i(b), i(c)$ indicate the three $\Im Z$ directions along the three tori.
So if one starts from Type I R-R field
\be
T_{i(a) i(b) i(c)} \to T_{23}, \qquad
T_{r(a) r(b) r(c)} \to T_{23 r(a) r(b) r(c) i(a) i(b) i(c)} = T_{01}
\ee
where $r(a), r(b), r(c)$ denote the three $\Re Z$ directions along the three tori,
and in the last equality one has used the fact that because of GSO projection
$S$ is chiral w.r.t.\ 10d tangent space Lorentz group:
\be
\Gamma_{0123 r(a) r(b) r(c) i(a) i(b) i(c)} S = S
\ee
So in the simplest version there are the vertex operators $T_{01}$ and $T_{23}$,
precisely the operators that we'll use in our computation.
Furthermore we note that the original D9-O9 system of type I theory,
upon 5 T dualities become D4-O4 system along $\sigma$ fixed point set namely $(0,1,r(1),r(2),r(3))$ directions.

When the orbifold group is of even order,
the original Type I theory compactified on $\IX$ already comes with O5 and D5-branes,
e.g.\ in the orbifold we are considering (namely Gimon-Polchinsky model \cite{Gimon:1996rq})
there are D5-branes along $(0,1,2,3, r(a), i(a))$ $(a=1,2,3)$
i.e.\ D5-branes wrapped on one of the $T^2$s.
These could be thought of as turning on some instantons on the remaining two $T^2$s,
modded by orbifold group, in the D9-branes.
Upon 5 T-dualities, for example for $a=1$, this will become D4-brane along
\be
(0,1, r(1), i(2),i(3))
\ee
Thus there will be several different species of D4 branes (and corresponding O4 planes).
Open string connecting any two species of D4 branes
have 4 directions that have mixed boundary conditions
i.e.\ Neumann at one end and Dirichlet at the other end.
On unoriented Riemann surfaces with boundaries and crosscaps,
we must allow all these different types of D4 and O4 systems,
however in the following, we will focus on the $\sigma$-fixed D4-O4 system,
as this is what survives in the smooth Calabi-Yau case that will be discussed in \cref{General CY},
even though these more general boundary conditions can also be dealt with by the same methods,
giving rise to definition of the real topological string
for this more general setting.

\subsection{Supergravity}
\label{sugra}

With the orientifold projection, supersymmetry is reduced to (2,2) in 2 dimensions,
namely 4 supercharges,
and we want to understand how to decompose $\NN=2$ chiral Weyl tensor\cite{Bergshoeff:1980is}
\be
\mathcal W^{ij}_{\mu\nu} = T^{ij}_{\mu\nu} - R_{\mu\nu\rho\lambda} \theta^i \sigma^{\rho\lambda} \theta^j + \cdots
\ee
which is anti-symmetric in $SU(2)$ indices $i,j\in\{1,2\}$ and anti-self-dual in Lorentz indices $\mu,\nu$
(namely, we will restrict both the graviphoton field strength $T$ and the Riemann tensor $R$,
which are antisymmetric in $\mu,\nu$, to a constant anti-self-dual background),
so that we may integrate some of its parts in 2d (2,2) superspace:
\be
\int d^4 x \int d^4 \theta \, \delta^2(\theta) \delta^2(x)
\left( \mathcal G_{\chi} - \frac12 \mathcal F_{g_\chi} \right)
\mathcal W_\parallel ^{g'}
\ee
where we removed from $\mathcal G$ in \cref{total-g-chi} the purely closed oriented piece;
let's denote such combination $\mathcal H_{g'}$, where $\chi=1-g'$.

We now discuss the term $\mathcal W_\parallel$.
Since orientifold action (in Weyl indices) takes the form
$\theta^i_\alpha \to 2 (\sigma^2)^i_{~j} (\sigma^{23})_\alpha^{~\beta} \theta^j_\beta$,
we can form orientifold even and odd combinations
\be
\theta^\pm = (\theta^1_1 \mp i \theta^2_2), \quad \tilde \theta^\pm = (\theta^2_1 \pm i \theta^1_2)
\ee
where $\theta^+$ and $\tilde{\theta}^+$ are the super-space coordinates
that are invariant under the orientifold.
For the Riemann sector, we write $\theta^1 \sigma^{\mu\nu} \theta^2$ as
\be
\begin{aligned}
\theta^1 \sigma^{23} \theta^2 = -\frac14 (\theta^+\tilde\theta^+ +\theta^-\tilde\theta^-) \\
\theta^1 \sigma^{02} \theta^2 = -\frac14 (\theta^+\tilde\theta^- +\theta^-\tilde\theta^+) \\
\theta^1 \sigma^{03} \theta^2 = -\frac14 (\theta^+\theta^- +\tilde\theta^+\tilde\theta^-)
\end{aligned}
\ee
This means that for example $\mathcal W_{01}$ decomposes as
\be
\mathcal W_{01}^{12} = T_{01}^{12} +\frac14 R_{0101} (\theta^+\tilde\theta^+ +\theta^-\tilde\theta^-) + \text{odd part}
\ee
The relevant (2,2) superfield is therefore
\be
\mathcal W_\parallel = T_{01}+ R_{0101} \theta^+ \tilde{\theta}^+ +\cdots
\ee
where $T_{01}$ etc.\ are  short hand notation for the anti-self-dual combination $T_{01}+T_{23}$.
From the original vector moduli super-fields $\Phi^a$ where $a$ labels different vector multiplets,
one gets (2,2) superfields
\be
 \Phi^a = \phi^a + F^a_{01} \theta^+ \tilde{\theta}^+ +\cdots
\ee
The effective action term that we are interested in is
\be
\int d\theta^+ d\tilde{\theta}^+  \mathcal H_{g'}(\Phi) \mathcal W_\parallel ^{g'} =
 \mathcal H_{g'}(\phi) R_{0101} T_{01}^{g'-1} +  D_a \mathcal H_{g'}(\phi) F^a_{01} T_{01}^{g'} + \cdots
\ee
where $D_a$ is the holomorphic covariant derivative w.r.t.\ $\phi^a$.
We will focus on the second term above,
as its computation is not affected by the issue mentioned in \cref{what-compute}.

\subsection{Vertex operators}
\label{inv-vert}

Recalling that this model is obtained by applying five T-dualities on Type I,
left and right moving sectors have opposite GSO projection.
We bosonize $(\psi^i,\tilde{\psi^i}) = (e^{i \phi_i},e^{i \tilde{\phi_i}})$,
so that orientifolding exchanges $\phi_i \leftrightarrow -\tilde{\phi_i}$;
in terms of bosonized fields
the 4-dimensional chiral spin fields with helicities labeled by $1$ and $2$ are
\be
S_1= e^{\frac{i}{2}(\phi_1+\phi_2)},
\quad S_2= e^{-\frac{i}{2}(\phi_1+\phi_2)},
\quad \tilde{S}_1= e^{\frac{i}{2}(\tilde{\phi}_1+\tilde{\phi}_2)},
\quad \tilde{S}_2= e^{-\frac{i}{2}(\tilde{\phi}_1+\tilde{\phi}_2)}
\ee
The internal spin fields are given by
\be
\Sigma= e^{\frac{i}{2}H}, \quad \bar{\Sigma} = e^{-\frac{i}{2}H},
\quad \tilde{\Sigma}= e^{\frac{i}{2}\tilde{H}}, \quad \tilde{\bar{\Sigma}} = e^{-\frac{i}{2}\tilde{H}}
\ee
where $H= \phi_3+\phi_4+\phi_5$ and $\tilde{H}=\tilde{\phi_3}+\tilde{\phi_4}+\tilde{\phi_5}$.
Finally, $\varphi$ and $\tilde{\varphi}$ appear in the bosonization of left and right moving superghosts.

The $(-1/2,-1/2)$ ghost picture for graviphoton vertex
\be
V_T^{(-1/2)}(p,\epsilon) =
p_\nu \epsilon_\mu : e^{-\frac12 (\varphi + \tilde \varphi)}
\left[S^\alpha (\sigma^{\mu\nu})^{~\beta}_\alpha  \tilde S_\beta \Sigma(z,\bar z) +
S_{\dot\alpha} (\bar \sigma^{\mu\nu} )^{\dot \alpha} _{~\dot\beta} \tilde S^{\dot\beta} \bar \Sigma(z,\bar z)\right]
e^{ip\cdot X} :
\ee
becomes for $T_{01}$ to the lowest order in momentum
\be
\label{graviph-vert}
 V_T =e^{-\frac{\varphi}{2}} e^{-\frac{\tilde{\varphi}}{2}}
( S_1\Sigma \tilde{S}_1\tilde{\bar{\Sigma} }- S_2\Sigma  \tilde{S}_2 \tilde{\bar{\Sigma}} )
\ee
The graviphoton vertex \cref{graviph-vert} is invariant under both the 
orbifold group as well as under the orientifold projection,
since orientifold action preserves the
$T_{01} (S_2 \tilde S_2-S_1 \tilde S_1)$
part of graviphoton vertex.

We take, for concreteness, the vector multiplet $\Phi^a$ to be the one corresponding to the K\"ahler modulus of the first torus
(i.e.\ in $Z^3$ direction).
In the $(-1,-1)$ picture
this vertex operator is just the (chiral, anti-chiral) primary $\psi^3 \tilde{\bar{\psi^3}}$.
The gauge field strength for this vector multiplet can be obtained by applying 
two supersymmetry transformations,
whose charges are obtained by integrating currents
$Q_\alpha = \oint dz \, j_\alpha$,
\be
j_\alpha = e^{-\frac \varphi 2} S_\alpha \Sigma,
\quad j_{\dot\alpha} = e^{-\frac \varphi 2} S_{\dot\alpha} \bar\Sigma,
\quad \tilde j_\alpha = e^{-\frac {\tilde \varphi} 2} \tilde S_\alpha \tilde\Sigma,
\quad \tilde j_{\dot\alpha} = e^{-\frac {\tilde\varphi} 2} \tilde S_{\dot\alpha} \bar{\tilde\Sigma}
\ee
and we recall that the orientifold only preserves 4 supercharges
\be
Q = Q_1 + \tilde Q_2, \quad Q' = Q_2 + \tilde Q_1,
\quad
\dot Q = Q_{\dot1} - \tilde Q_{\dot2}, \quad \dot Q' = Q_{\dot2} - \tilde Q_{\dot1}
\ee
The result for an orbifold group and orientifold action invariant such vertex is
\be
 V_{F^a}= e^{-\frac{\varphi}{2}}e^{-\frac{\tilde{\varphi}}{2}}(S_1 e^{\frac{i}{2}(\phi_3-\phi_4-\phi_5)} \tilde{S}_1 
 e^{\frac{i}{2}(-\tilde{\phi}_3+\tilde{\phi}_4+\tilde{\phi}_5)}) -\{(S_1, \tilde{S}_1 )\rightarrow (S_2, \tilde{S}_2 )\}
\ee

\subsection{Computation}
\label{Computation}

We are interested in computing the amplitude involving $g'$ $V_T$ and one $V_{F^a}$ on a surface $\Sigma_{g,h,c}$,
where we recall that $(g,h,c)$ denote respectively the number of handles, boundaries and crosscaps
and $g'$ is the genus of the double cover of $\Sigma_{g,h,c}$, namely $g'=2g+h+c-1$.
By using the image method this amplitude can be computed on the compact oriented Riemann surface $\Sigma_{g'}$.
Denoting the image of a point $p \in \Sigma_{g,h,c}$ as $\bar{p} \in \Sigma_{g'}$
and using the fact that the right moving part of the vertex at $p$ is mapped to a left moving part
dictated by the orientifold action at $\bar{p}$,\footnote
{Note that the image across a boundary is governed by the Neumann or Dirichlet boundary conditions,
with Dirichlet directions being accompanied by $\IZ_2$ involution.
The resulting $\IZ_2$ involution is the same that appears with orientifolding action.
This is so because the D4 and O4 planes are parallel.
Had we considered a situation where they were not parallel or a system containing say D0-branes with O4-planes,
these two involutions would have been different and we would have to go to
quadruple covers for world-sheets containing both boundaries and crosscaps.}
we get
\be
V_T(p)|_{p \in \Sigma_{g,h,c}}
&= e^{-\frac{\varphi}{2}}(p) e^{-\frac{\varphi}{2}}(\bar{p})(S_1(p) \Sigma(p) 
S_2(\bar{p})\Sigma(\bar{p}) -  S_2(p) \Sigma(p) S_1(\bar{p})\Sigma(\bar{p}))|_{p \in \Sigma_{g,h,c}} \\
&= e^{-\frac{\varphi}{2}}(p) e^{-\frac{\varphi}{2}}(\bar{p}) S_1(p) \Sigma(p) S_2(\bar{p})\Sigma(\bar{p})|_{p\in \Sigma_{g'}}
\ee
Similarly
\be
V_{F^a}|_{p \in \Sigma_{g,h,c}}
&= e^{-\frac{\varphi}{2}}(p) e^{-\frac{\varphi}{2}}(\bar{p}) 
(S_1(p) \hat{\Sigma}(p) S_2(\bar{p})\hat{\Sigma}(\bar{p}) -  
S_2(p) \hat{\Sigma}(p) S_1(\bar{p})\hat{\Sigma}(\bar{p}))|_{p \in \Sigma_{g,h,c}}\\
&= e^{-\frac{\varphi}{2}}(p) e^{-\frac{\varphi}{2}}(\bar{p})S_1(p) \hat{\Sigma}(p)
S_2(\bar{p})\hat{\Sigma}(\bar{p})|_{p\in \Sigma_{g'}}
\ee
where $\hat{\Sigma}=e^{\frac{i}{2}(\phi_3-\phi_4-\phi_5)}$.
In other words both for $V_T$ and $V_{F^a}$ the region of integration extends to double cover $\Sigma_{g'}$.
This is actually due to the fact that both these operators are orientifold invariant.
Note that the graviphoton as well as matter field strength vertices in $(-1/2,-1/2)$ picture
already come with one momentum giving altogether $(g'+1)$ momenta.
Therefore in the remaining part of the vertices we can set zero momenta.
Finally we can write the amplitude of interest on the double cover as
\be
A &=\int_{\Sigma_{g,h,c}} d^2 z \prod_{i=1}^{g'} d^2 x_i
\langle V_{F_a}(z) \prod_{i=1}^{g'} V_T(x_i) \rangle_{\Sigma_{g,h,c}}\\
&=\int_{\Sigma_{g'}} d^2 z \prod_{i=1}^{g'} d^2 x_i 
\langle e^{-\frac{\varphi}{2}}(z) e^{-\frac{\varphi}{2}}(\bar{z}) S_1(z) \hat{\Sigma}(z) S_2(\bar{z})\hat{\Sigma} (\bar{z})\\
& \quad \times \prod_{i=1}^{g'}e^{-\frac{\varphi}{2}}(x_i) e^{-\frac{\varphi}{2}}(\bar{x}_i)
S_1(x_i) \Sigma(x_i) S_2(\bar{x}_i) \Sigma(\bar{x}_i) 
\prod_{a=1}^{3 g'-1} \pco (u_a) \prod_{a=1}^{3g'-3} \int \mu_a b \rangle
\ee
where the number $(3 g'-1)$ of PCOs follows from the fact that the total picture (super-ghost charge)
on a genus $g'$ surface must be $2g'-2$ and
the operators that are inserted give a total super ghost charge $-(g'+1)$.
As each $\Sigma$ operator carries internal $U(1)$ charge $3/2$ and $\hat{\Sigma}$ carries a charge $-1/2$,
the total internal $U(1)$ charge carried by the vertices is $3/2 (2 g')-1/2(2) = 3 g'-1$.
This means that to balance the total $U(1)$ charge, each of the PCO must contribute an internal charge $(-1)$.
Thus the relevant part of each PCO is $e^{\varphi} G_-$,
where $G_-$ is the supercurrent of the $\NN=2$ superconformal field theory describing the Calabi-Yau space
(the subscript $-$ refers to the $U(1)$ charge).
In the orbifold example we are considering, fermion charge for each plane must be conserved,
which implies that of the $(3 g'-1)$ PCO,
$(g'+1)$ contribute $e^{\varphi} \bar{\psi^3} \partial Z^3$
and $(g'-1)$ each contribute $e^{\varphi} \bar{\psi^4} \partial Z^4$ and $e^{\varphi} \bar{\psi^5} \partial Z^5$ respectively.
We shall denote the positions of these three groups of PCOs by
$u^{(1)}_a$ for $a=1,\ldots,g'+1$ and $u^{(2)}_a$ and $u^{(3)}_a$ for $a=1,\ldots,g'-1$.
Of course we will need to sum over all the partitions with appropriate anti-symmetrization.
Finally $\mu_a$ for $a=1,\ldots,3g'-3$ are the Beltrami differentials and $b$ are anti-commuting spin 2 ghost fields.
Note that $b$ provide the $3g'-3$ quadratic differentials $h_a$ that are dual to the Beltrami differentials $\mu_a$.

We use chiral bosonization formulae for anti-commuting $(b,c)$ system with conformal dimensions $(\lambda, 1-\lambda)$
with $\lambda >1$ and $g'>1$\cite
{Verlinde:1987sd,Verlinde:1986kw,Ohmori:2013zla,Sen:2015hia,Lechtenfeld:1989wu,Morozov:1989ma},
namely
\be
 \langle \prod_{i=1}^{Q(g'-1)} b(x_i)\rangle =
\mathcal Z_1^{-\frac{1}{2}} \theta(\sum_i x_i - Q \Delta) \prod_{i<j}E(x_i,x_j) \prod_i\sigma^Q(x_i)
\ee
where $Q=2\lambda -1$ and by Riemann-Roch theorem the number of zero modes of $\lambda$-differential $b$ is $Q(g'-1)$
and the $(1-\lambda)$-differential $c$ has no zero mode for $g'>1$.
$\mathcal Z_1$ is the chiral non-zero determinant of the Laplacian acting on a scalar.
The above expression is also valid for the case of $\lambda=1$ when the $(b,c)$ system is twisted.
In particular, for spacetime and internal fermions the functional integral
can be expressed in terms of chiral fermionic $(b,c)$ theories with general spin $\lambda$ respectively $1-\lambda$:
using bosonization the following expression was found for the functional integral
in spin structure $s$ and with $\sum q_i =Q(g'-1)$
\be
\int [Db Dc]_s \exp (-S[b,c]) \, \prod b(z_i) \prod c(w_j)
=Z[s] \left( \sum z_i -\sum w_j -(2\lambda-1)\Delta \right),\\
Z[s]\left( \sum q_i z_i -Q \Delta \right) = \mathcal Z_1^{-\frac12} \theta_s\left( \sum q_i z_i -Q \Delta \right)
\prod_{i<j} E(z_i,z_j)^{q_i q_j} \prod_i \sigma(z_i)^{q_i Q}
\ee
For the super ghost correlation functions, we use
\be
\langle \prod^{n+1} \xi (x_i) \prod^n \eta (y_j) \prod \exp [q_k \varphi (z_k)] \rangle_s =\\
\frac{\prod_{j=1}^n \theta_s \left(- y_j + \sum x - \sum y + \sum q z - 2 \Delta \right)}
{\prod_{i=1}^{n+1} \theta_s \left(- x_i + \sum x - \sum y + \sum q z - 2 \Delta \right)}
\frac{\prod_{i<i'} E(x_i,x_{i'}) \prod_{j<j'} E(y_j,y_{j'})}
{\prod_{i,j} E(x_i,y_j) \prod_{k<l} E(z_k,z_l)^{q_k q_l} \prod_k \sigma(z_k)^{2q_k}}
\ee
where in our case $n=0$ and one $\xi$ is implicitly there;
moreover, the overall minus sign in the denominator $\theta$ argument that we will have
can be justified by GSO projection.\footnote
{Namely, one should sum over spin structures with $\sum_s \varepsilon_s f(s)$,
where $\varepsilon$ is a sign and $f$ the relevant function,
and then require monodromy invariance:
this translates for our case in a sign shift for odd spin structure, which gives the result below.}
Some relevant facts about $\theta$ functions are summarized in \cref{theta}.
We can write the correlation function of interest as
\be
A &= K ~ \sum_{s}
\frac
{\theta_s(\frac{1}{2}(\sum_i(x_i-\bar{x}_i)+z-\bar{z}))^2
\theta_{s,g_1}(\frac{1}{2}(\sum_i(x_i+\bar{x}_i)+z+\bar{z})-\sum_a u^{(1)}_a) }
{\theta_s(\frac{1}{2}(\sum_i(x_i+\bar{x}_i)+z+\bar{z})- \sum_a u_a + 2 \Delta)}\\
& \quad \times \prod_{k=2}^3\theta_{s,g_k}(\frac{1}{2}(\sum_i(x_i+\bar{x}_i)-z-\bar{z})-\sum_a u^{(k)}_a)
\label{A}
\ee
where the sums appearing above are in the appropriate ranges,
for example for $x_i$, $i=1,\ldots,g'$, for $u^{(1)}_a$, $a=1,\ldots,g'+1$,
for $u^{(2)}_a$ and $u^{(3)}_a$, $a=1,\ldots,g'-1$ and finally $a=1,\ldots,3g'-1$ for $u_a$.
The twists $g_1,g_2,g_3$ are the orbifold twists along the three tori
along $2 g'$ cycles (in other words $g_i$ are points in Jacobi variety).
Since the orbifold group $G \subset SU(3)$ in order to preserve supersymmetry,
we have the relation $g_1+g_2+g_3=0$.
Finally $K$ is the spin-structure independent part of correlation function
and can be expressed in terms of prime forms and certain nowhere vanishing holomorphic sections $\sigma$ that are
quasi-periodic and transform as $\frac{g'}2$ differential under local coordinate transformations.
The prime form $E(x,y)$ has the important property that it vanishes only at $x=y$ in $\Sigma_{g'}$,
transforms as holomorphic $-\frac12$ differentials in arguments $x$ and $y$
and is quasi-periodic along various cycles.
In fact $K$ can be determined by just the leading singularity structures coming from OPE
and the total conformal weights at each point.
The position dependent part of $K$ is given by
\be
\label{K}
K&= \frac{\prod_{i<j} E(x_i,x_j) E(\bar{x}_i,\bar{x}_j)} {\prod_i E(x_i,\bar{z})E(\bar{x}_i,z)}
\frac{\prod_{k=2}^{3} \prod_{a=1}^{g'-1} E(z,u^{(k)}_a) E(\bar{z},u^{(k)}_a)} {\prod_{k<l}\prod_{a,b} E(u^{(k)}_a,u^{(l)}_b)}
\frac{\sigma(z)\sigma(\bar{z}) \prod_i \sigma(x_i)\sigma(\bar{x}_i)} {\prod_a \sigma(u_a)^2}\\
& \quad \times \mathcal Z_1^{-2}
\langle \prod \partial Z^3(u^{(1)}) \partial Z^4(u^{(2)}) \partial Z^5(u^{(3)}) \prod_k \int \mu_k b \rangle
\ee
where $\langle \cdots \rangle$ indicates correlation function
in the space of all bosonic fields $(Z^1,\ldots,Z^5)$ and the $(b,c)$ ghost system.
We can now put two of the PCO say at $u_{3g'-1}$ and $u_{3g'-2}$ at points $z$ and $\bar{z}$ respectively.
The expression for $K$ above shows that if these two PCO positions appear
in the partitioning as $u^{(2)}$ or $u^{(3)}$
the result vanishes due to the appearance of $\prod_{k=2}^{3} \prod_{a=1}^{g'-1} E(z,u^{(k)}_a) E(\bar{z},u^{(k)}_a)$ in the numerator.
What this means is that the only non-vanishing contribution can come when these two PCOs are
in the partitioning $u^{(1)}$.
There are now $g'-1$ remaining $u^{(1)}$, the same number as the ones for $u^{(2)}$ and $u^{(3)}$.
The amplitude now becomes much simpler:
\be
\label{set-pco-z}
A=K~ \sum_{s}\frac
{\theta_s(\frac{1}{2}(\sum_i(x_i-\bar{x}_i)+z-\bar{z}))^2
\prod_{k=1}^3 \theta_{s,g_k}(\frac{1}{2}(\sum_i(x_i+\bar{x}_i)-z-\bar{z})-\sum_a u^{(k)}_a)}
{\theta_s(\frac{1}{2}(\sum_i(x_i+\bar{x}_i)-z-\bar{z})-\sum_a u_a + 2 \Delta)} 
\ee
where the position dependent part of $K$ is
\be
K&=
\frac{\prod_{i<j} E(x_i,x_j) E(\bar{x}_i,\bar{x}_j)} {\prod_i (E(x_i,\bar{z}) E(\bar{x}_i,z)}
\frac1 {\prod_{1\leq i<j \leq 3} E(u^{(i)},u^{(j)})}
\frac{\prod_i \sigma(x_i)\sigma(\bar{x}_i)} {\sigma(z)\sigma(\bar{z})\prod_a \sigma(u_a)^2}\\
& \quad \times
\mathcal Z_1^{-2}
\langle
\partial Z^3(z) \partial Z^3(\bar{z})
\prod_{n=1}^3 \partial Z^{n+2} (u^{(n)})
\rangle
\langle \prod_a \int \mu_a b \rangle
\ee
We can now choose the following gauge condition so that
superghost theta function appearing in the denominator cancels with one of the spacetime theta functions:
\be
 \sum_{a=1}^{3g'-3} u_a = \sum_{i=1}^{g'} \bar{x}_i -z + 2 \Delta
\ee
After performing the spin-structure sum using \cref{riem-id}, the result is
\be
 A= K~ \theta(\sum_i x_i - \bar{z}-\Delta) \prod_{k=1}^3 \theta_{g_k}( \Delta-\sum_a u^{(k)}_a )
\ee
We now multiply this expression by the identity
\be
 1= \frac{\theta(\sum_i \bar{x}_i - z-\Delta)}{\theta(\sum_a u_a -3 \Delta)}
\ee
which follows from the gauge condition, and make use of chiral bosonization formulae
\be
\label{bosoniz}
\mathcal Z_1^{-\frac{1}{2}} \theta_{g_k}(\Delta-\sum_{a=1}^{g'-1} u^{(k)}_a)
\prod_{a<b} E(u^{(k)}_a,u^{(k)}_b) \prod_a \sigma(u^{(k)}_a)
&=
\langle \prod_a \bar{\psi^k} (u^{(k)}_a)\rangle\\
\mathcal Z_1^{-\frac{1}{2}} \theta(\sum_{i=1}^{g'} x_i - \bar{z}-\Delta)
\frac{\prod_{i<j} E(x_i,x_j)}{\prod_i E(x_i,\bar z)}
\frac{\prod_i \sigma(x_i)}{\sigma(\bar z)}
&=
\langle \prod_i\bar{\psi}(x_i)\psi(\bar{z})\rangle\\
&=
\mathcal Z_1 \det (\omega_i(x_j))\\
\mathcal Z_1 ^{-\frac{1}{2}} \theta(\sum_{a=1}^{3g'-3} u_a -3 \Delta)
\prod_{a<b} E(u_a,u_b) \prod_a \sigma(u_a)^3
&=
\langle\prod_a b(u_a)\rangle
\ee
where $(\bar{\psi^k},\psi^k)$ are anti-commuting $(1,0)$ system twisted by $g_k$,
$\omega_i$ are the $g'$ abelian differentials and $(b,c)$ anti-commuting spin-$(2,-1)$ system.
Note that by Riemann-Roch theorem $b$ has $(3g'-3)$ zero modes (quadratic differentials) and therefore
the last correlation function just soaks these zero modes.
It is interesting to note that after the spin structure sum,
we have obtained the correlation function
in topologically twisted internal theory
where $\bar{\psi^k}$ and $\psi^k$ are of dimension $(1,0)$ for $k=3,4,5$.
Taking into account various $\partial Z^k$ in $K$ and summing over all partitions we obtain
\be
A &= \int_{\Sigma_{g'}} d^2 z \prod_i d^2 x_i (\det \omega(x_i)) (\det \omega(\bar{x}_i)) \mathcal Z_1^2\\
& \quad \times \frac
{\langle \partial Z^3(z) \partial Z^3(\bar{z}) \prod_{a=1}^{3g'-3} G_-(u_a) \prod_{a=1}^{3g'-3} \int(\mu_a b) \rangle}
{\langle \prod_{a=1}^{3g'-3} b(u_a)\rangle}
\ee
where 
$G_-= \sum_{k=3}^5 \bar{\psi^k} \partial Z^k$ is the twisted super current of dimension 2 and
$G_+=\sum_{k=3}^5 \psi^k \partial \bar{Z^k}$ is dimension 1 topological BRST current.
$G_-$ in $A$ provide only zero modes both for $\bar{\psi^k}$ and $\partial Z^k$
and hence must be holomorphic quadratic differentials.
Therefore $\frac{\prod_{a=1}^{3g'-3} G_-(u_a)}{\langle \prod_{a=1}^{3g'-3} b(u_a)\rangle}$ is independent of $u_a$.
Finally, using the fact that the anti-analytic $\IZ_2$ involution $\Omega$ maps abelian differentials
as $\omega_i(\bar{x}) = \sum_j \Gamma^\Omega_{i j} \bar{\omega}_j(x)$,
where $\Gamma:=\Gamma^\Omega$ is a matrix satisfying $\Gamma^2 =I$, we find that
\be
\prod_i d^2 x_i (\det \omega_j(x_k)) (\det \omega_j(\bar{x}_k))
&=
\prod_i d^2 x_i (\det \omega_j(x_k)) (\det (\Gamma \,\bar{\omega})_j(x_k))\\
&\sim
\det \Gamma \,\det \Im \tau
\ee

The amplitude then becomes
\be
\label{holo-deriv}
A
&= \int_{\mathcal M_{g,h,c}} \det \Gamma \, \det \Im \tau ~ \mathcal Z_1^2
\langle \int d^2 z \partial Z^3(z) \partial Z^3(\bar{z}) \prod_a (\int \mu_a G_-)\rangle\\
&= D_{t_3} \int_{\mathcal M_{g,h,c}} \det \Gamma ~ (\det \Im \tau)~\mathcal Z_1^2 \langle \prod_a (\int \mu_a G_-)\rangle
\ee
where we have used the fact that
$\int_{\Sigma_{g'}} d^2 z \partial Z^3(z) \partial Z^3(\bar{z})=
2 \int_{\Sigma_{g,h,c}} d^2 z \partial Z^3(z) \bar{\partial Z^3}(z)$
is the marginal operator corresponding to the complexified K\"ahler modulus $t_3$ of the torus along $Z^3$ direction
and hence gives a holomorphic covariant derivative with respect to $t_3$.
The integral is over $\mathcal M_{g,h,c}$
since among the Beltrami on $\Sigma_{g'}$
we should only include those that are invariant under the involution.\footnote
{On $\Sigma_{g'}$ there are a total of $6g'-6$ real moduli,
but the fact that we are actually restricting to $\Sigma_{g,h,c}$ reduces these to $3g'-3$ real moduli.}
Furthermore $\langle \cdots \rangle$ denotes correlation function in the space of bosonic fields $(Z^1,Z^2)$
and the topologically twisted $\NN=2$ superconformal theory describing the Calabi-Yau manifold.

Finally $\mathcal Z_1^2 \det \Im \tau$ cancels with the partition function of the spacetime bosonic fields $(Z^1,Z^2)$.
This can be seen using the results of\cite{Blau:1987pn}, as follows:

\begin{enumerate}

\item Consider first diagrams that have no crosscaps.
Let $\Sigma_{g'}$ be the double cover of $\Sigma_{g,h}$ and
let $\Omega$ be the map that takes a point in $\Sigma_{g,h}$ to its image in $\Sigma_{g'}$.
Then for a Neumann (N) direction $X_N$ the scalar determinant (denoted by $\det\Delta^+$)
is over functions that are even under $\Omega$
while the determinant for Dirichlet (D) directions $X_D$ (denoted by $\det\Delta^-$) is over odd functions under $\Omega$.
These are given in\cite[eqs.~(4.1-4.3)]{Blau:1987pn}.
Note that correction factor denoted by $R$ appears with opposite powers in the two cases.
This means that if one has equal number of N and D directions (in our case number of N and D is 2 each),
then the correction factor cancels and one just gets the square root of the closed string determinants for 4 scalars.
The closed string result for this is $1/(\det \Im \tau)^{4/2}$.
The square-root of this gives $1/(\det \Im \tau)$.
So for this to work it is crucial that D-branes have equal number of N and D $\IR^4$ directions.

\item Now consider diagrams with just crosscaps (i.e.\ no boundaries).
The formula given in\cite[eq.~(4.25)]{Blau:1987pn} is just for even functions,
but this is because in\cite{Blau:1987pn} orientifolding is simply worldsheet parity operator.
In our case however it is combined with a $\IZ_2$ reflection of the two D-directions of $\IR^4$.
This means that if $p \in \Sigma_{g'}$ is mapped to $\bar{p}$ under $\Omega$,
then $X_N(p) = X_N(\bar{p})$ while $X_D(p)= - X_D(\bar{p})$.
So once again for the two cases one has $\det\Delta^+$ and $\det\Delta^-$ respectively
and using\cite[eq.~(4.3)]{Blau:1987pn} again correction factors cancel.

\item The reason why in\cite{Blau:1987pn} the authors needed to look at quadruple cover
for the surfaces that have both crosscaps and boundaries, is as follows.
One goes to the quadruple cover as explained in the third paragraph of page 287:
one first goes to an unoriented boundary-less double cover $B$ of $\Sigma_{g,h,c} = B/\Omega_1$,
but $B$ still has the crosscaps inherited from $\Sigma_{g,h,c}$.
One now goes to oriented double cover $Q$ of $B= Q/\Omega_2$ so that
the original $\Sigma_{g,h,c} = Q/(\IZ_2 \times \IZ_2)$, where the two $\IZ_2$ are generated by $\Omega_1$ and $\Omega_2$.
They need to work with $Q$ (as is seen in the table eq.~(4.24))
because for them $\Omega_1$ and $\Omega_2$ act differently on $X_D$:
$X_N(p)=X_N(\Omega_1(p))=X_N(\Omega_2(p))$ but $X_D(p)=-X_D(\Omega_1(p))=X_D(\Omega_2(p))$.

However for us, since orientifold action comes with $\IZ_2$ reflection on $X_D$:
$X_D(p)=-X_D(\Omega_1(p))=- X_D(\Omega_2(p))$;
therefore one can just work with the complex double described in third paragraph of page 287
and $X_N$ and $X_D$ will be even and odd functions under $\Omega$ and once again the correction factors cancels.

\end{enumerate}

So it is crucial not only that there are two N and two D-directions in $\IR^4$
but also that orientifold action comes with $\IZ_2$ reflection on $X_D$,
otherwise the prefactors would not have canceled by integrating the positions of the vertices.
This is precisely what happens in our case, as we have one D4-brane (extended along two spacetime directions)
stuck on top of the O4-plane.

Thus the amplitude \cref{holo-deriv} for a fixed genus of the covering space $g'$ reduces to
\be
\label{finalresult}
A_{g'} =  D_{t_3} \sum \int_{\mathcal M_{g,h,c}} \det \Gamma_{g,h,c}
\langle \prod_a (\int \mu_a G_-)\rangle_ \text{twisted internal theory}
\ee
where the sum is over all $(g,h,c)$ such that $g'=2g+h+c-1$ and
$\Gamma_{g,h,c}$ is the corresponding involution in $\Sigma_{g'}$.

Even though in the preceding computation we have implicitly used only the $\sigma$ fixed D4 and O4 systems
at the boundaries and the crosscaps in $\Sigma_{g,h,c}$,
we could have allowed the other D4-O4 systems as well.
The crucial point is that all of these D4-O4 systems have the same boundary conditions on $\IR^4$
(they differ only along the orbifold directions).
This means that the cancellation of the contributions of $\IR^4$ bosons and fermions
as well as $(b,c), (\beta,\gamma)$ systems
continues to hold and one again ends up with the partition function of the twisted internal theory on $\Sigma_{g,h,c}$,
now allowing more general boundary conditions relevant to different species of D4-O4 systems
that appear in Gimon-Polchinsky like models that satisfy local tadpole cancellation condition.
It will be interesting to study these more general real topological strings.

\subsection{General Calabi-Yau case}
\label{General CY}

For generic CY, we can extend the above analysis using CFT arguments as in \cite{Antoniadis:1993ze}.
The idea is to use the CFT description of CY in terms of $\NN=2$ SCFT,
and the fact that the graviphoton vertices can be expressed in terms of the scalar field $H$ and $\tilde{H}$
that bosonize the left and right moving $U(1)$ currents $J$ and $\tilde{J}$
of the $\NN=2$ SCFT as $J = i\sqrt{3} \partial H$ and $\tilde{J} = i \sqrt{3} \partial \tilde{H}$.
The internal part of the graviphoton vertices are
$\Sigma \tilde{\bar{\Sigma}}$ where $\Sigma = 
e^{i \frac{\sqrt{3}}{2} H}$ and $\tilde{\bar{\Sigma}}=e^{-i \frac{\sqrt{3}}{2} \tilde{H}}$.
Similarly the internal part of the vertex $V_{F^a}$ for a vector multiplet (K\"ahler modulus) field strength
is obtained by the spectral flow of (chiral, antichiral) operator and is given by
$\hat{V}\hat{\tilde{V}} e^{-\frac{i}{2\sqrt{3}} H} e^{\frac{i}{2\sqrt{3}} \tilde{H}}$
where $\hat{V}\hat{\tilde{V}}$ has dimension $(1/3,1/3)$ and has no singular OPE with $U(1)$ currents.\footnote
{It may be not always possible to decompose such operator in a product $\hat{V}\hat{\tilde{V}}$;
however, for our argument below, this issue does not matter, since whatever this decomposition is,
it will always have the behavior we describe under the image trick.}
The marginal operator that defines the corresponding K\"ahler modulus $t^a$ is
$V_{t^a} =\oint G_- \oint \tilde{G}_+ V_a$,
where $V_a=\hat{V}\hat{\tilde{V}} e^{\frac{i}{\sqrt{3}} H} e^{-\frac{i}{\sqrt{3}} \tilde{H}}$,
$G_- = e^{-\frac{i}{\sqrt{3}} H} \hat{G}_-$, $\hat{G}_-$ have dimensions $4/3$
and non-singular OPE with $U(1)$ current,
and similarly $\tilde{G}_+ =  e^{\frac{i}{\sqrt{3}} \tilde{H}} \hat{\tilde{G}}_+$.

The important point is that the entire spin structure dependence in the internal theory is encoded
in the $U(1)$ charge lattices.
As was shown in\cite{Lerche:1988np, Lechtenfeld:1989be, Lust:1988yf},
the characters of the $\NN=2$ SCFT together with one $SO(2)$ character of a free complex fermion
are given by the branching functions $F_{\Lambda,s}(\tau)$ of level 1 $E_6/SO(8)$ coset theory:
\be
\chi_{\Lambda}(\tau) = \sum_s F_{\Lambda,s}(\tau) \chi_s(\tau)
\ee
where $\chi_{\Lambda}$ are three level 1 $E_6$ characters
with $\Lambda$ labelling conjugacy classes $ (1), (27), (\bar{27})$
and $\chi_s$ are the four level 1 $SO(8)$ characters with $s$ labelling the four spin-structures.

The characters of the internal SCFT together with one complex fermion can then be expressed as
$F_{\Lambda,s}(\tau)\ch_{\lambda}(\tau)$
where $\ch_{\lambda}(\tau)$ represents the contribution of the rest of the internal CFT.
The essential point is that $\ch_{\lambda}(\tau)$ depends only on $\Lambda$
and not on the spin-structures.
This allows one to perform the spin structure sum without knowing the details of the internal SCFT.
The generalization to higher genus is obtained
by assigning an $E_6$ representation $\Lambda$ for each loop and including all the spin-structures.

Under the world sheet parity $H \leftrightarrow \tilde{H}$,
while under the anti-analytic involution that takes $(3,0)$ form to $(0,3)$ form,
$(H,\tilde{H}) \rightarrow (-H,-\tilde{H})$.
Under the combined action therefore $H \leftrightarrow -\tilde{H}$.
This means that in the double cover $\Sigma_{g'}$ of $\Sigma_{(g,h,c)}$,
using the image trick, the internal part of
$V_T(p) \rightarrow e^{i \frac{\sqrt{3}}{2} H}(p)e^{i \frac{\sqrt{3}}{2} H}(\bar{p})$
and similarly
$V_{F^a}(p) \rightarrow \hat{V}(p) e^{-\frac{i}{2\sqrt{3}} H}(p) \hat{\tilde{V}}(\bar{p})e^{-\frac{i}{2\sqrt{3}} H}(\bar{p})$.

We will not give the details of the computation here,
as the steps are very similar to that in \cite{Antoniadis:1993ze} apart from the insertion of $V_{F^a}$.
One can carry out the spin structure sums,
extract the moduli derivative w.r.t.~$t^a$
and integrate the positions of the graviphoton and graviton vertices resulting in \cref{finalresult}.

\subsection{Comment on \texorpdfstring{$g'=1$}{g'=1} case}
\label{g'=1}

In this case, a direct calculation of $\mathcal H_{g'}$ is not possible,
as on-shell one-point function of graviton vertex will vanish.
However the amplitude considered here that computes holomorphic moduli derivative of $\mathcal H_{g'}$
involves the two point function $\langle F_T F_a \rangle$,
which is not zero on-shell due to the fact that along $x^2,x^3$ directions the momenta are not conserved
because of the $\sigma$ involution.
One special feature for $g'=1$ is that there is one zero mode each for $b$ and $c$ ghost
(this basically means that there is only one real world-sheet modulus and
there is one real translational invariance that needs to be fixed).
In order to soak the ghost zero modes we need to insert one $b$ field that is folded with the Beltrami
and one $(c+ \tilde{c})$ field that can be put at say $z$.
As a result the operator at $z$ has total left plus right dimension one and
hence the vertex at $z$ is integrated on a line transversal to the translational symmetry orbit.

One can repeat the calculation in \cref{Computation} up to \cref{A,K}.
Now, however, there are only two PCOs and we cannot choose the gauge condition
where both the PCOs are put at $z$ and $\bar{z}$,
as this would not allow  cancellation of the superghost theta function.
Instead we can put the two PCOs at $\bar{x}$ and $\bar{z}$.
With this gauge choice, the superghost theta function cancels with one of the space-time fermion theta function
and the spin structure sum leads to
\be
A= \frac{\theta_1(x-\bar z) \theta_{1,g_1}(\bar x-z) \theta_{1,g_2}(0) \theta_{1,g_3}(0)}{E(x,\bar{z})E(\bar{x},z)}
\partial Z^3(\bar{x})\partial Z^3(\bar{z})
\ee
$\theta_1$ is the odd theta function and $\theta_{1,g_i}$ are odd theta functions twisted by $g_i$.
Recalling that $E(x,\bar{z})= \theta_1(x-\bar z)$ and the Szego kernel
\be
\frac{\theta_{1,g_1}(\bar x-z)}{E(\bar{x},z)} \sim \langle \bar{\psi^3}(\bar{x}) \psi^3(z) \rangle_\text{odd}
\ee
we can rewrite the amplitude as an amplitude in the twisted internal theory
(i.e.\ fermions are in odd spin structure twisted by the orbifold group)
\be
\langle G_-(\bar{x}) \psi^3(z) \partial Z^3(\bar{z}) \rangle
\ee
In the orbifold case $\langle \partial Z^3(\bar{x}) \partial Z^3(\bar{z}) \rangle$ will be zero
unless $\partial Z^3$ has zero modes.
This can happen only if $Z^3$ is untwisted i.e.\ $g_3 =0$.
In this case $\psi^3$ and $\bar{\psi^3}$ have zero modes and the above expression becomes
\be
A = D_{t_3} \mathcal{H}_1
\ee
where $\mathcal{H}_1$ is as given in \cite[Eq.(4.11)]{Walcher:2007qp}.
Note that the insertion of the fermion number current soaks the $\psi^3$ and $\bar{\psi^3}$ zero modes.
Thus in the orbifold theory the amplitude is non-vanishing only if the orbifold admits $\NN=4$ subsectors
and in that case it gives the derivative of $\mathcal{H}_1$ with respect to the K\"ahler modulus of the untwisted $T^2$.

In the general CY case, one can again carry out the analysis as described in the previous subsection
and the correlation function is
\be
\label{g1}
\int \frac{d\ell}{\ell^2} \int d^2 x \, d^2 z \, \langle (G_-+\tilde{G}_+)(x)\oint  (G_--\tilde{G}_+) V_a(z) \rangle_{\Sigma_{(0,h,c)}}
\ee
where $(h+c)=2$ and $V_a$ is the (chiral, antichiral) primary corresponding to the K\"ahler modulus $t^a$.
Note that $V_a$ carries charge $(+1,-1)$ with respect to left and right $U(1)$ currents.
Here $\ell$ is the evolution parameter for open string (for annulus and M\"obius) and for closed string (for Klein bottle)
and the factor $1/\ell^2$ can be understood as follows:
$1/\ell$ comes from dividing by the volume of the residual translational invariance
and another $1/\ell$ comes from integrating the spacetime momenta along $(x^0,x^1)$ directions
(note that since the $(x^2,x^3)$ are transverse to D4 and O4 planes, there are no momenta integrals along these directions).
This expression can be further simplified by writing $(G_-+\tilde{G}_+)(x) = \oint (G_-+\tilde{G}_+) (J- \tilde{J})(x)$
and deforming the contour 
\be
\int \frac{d\ell}{\ell^2} \int d^2 x \, d^2 z \, \langle (J- \tilde{J})(x) \oint G_- \oint \tilde{G}_+ V_a(z) \rangle_{\Sigma_{(0,h,c)}}
= D_{t^a} \mathcal{H}_1
\ee
where we have used the fact that $\oint G_- \oint \tilde{G}_+ V_a(z)$ is the marginal operator
corresponding to the deformation of the K\"ahler modulus $t^a$
and $\int d^2 x(J- \tilde{J})(x) = \ell F$ where $F$ is the left minus right $U(1)$ charge.

\section{Conclusions and open issues}
\label{conclusion}

In this paper we have shown that real topological string amplitudes can be obtained
as a subsector of physical type I superstring amplitudes
by extending the construction of \cite{Antoniadis:1993ze} to the presence of a D4/O4 system.
Having obtained Walcher's topological string in terms of physical type I amplitude,
a natural question is what its heterotic dual would be.
This is particularly useful to study the singularity structure of the topological string
when some massive state becomes massless as one moves in the moduli space of compactification.
In the type I or type II side such states are necessarily
some D-brane states wrapped on a vanishing cycle and hence non-perturbative,
but on the heterotic side one can realize the would be massless states perturbatively.
In fact in the context of standard $\mathcal F_g$ in the oriented type II theory,
a study on the heterotic dual\cite{Antoniadis:1995zn} gave the Schwinger formula
describing the singularity structure that explicitly proved the $c=1$ conjecture
and was generalized to all BPS states by Gopakumar and Vafa\cite{Gopakumar:1998ii,Gopakumar:1998jq}.
The singularity structure for the real topological string appears to be much more complicated 
involving both Bernoulli and Euler numbers \cite{Krefl:2010fm}.
It will be interesting to find a dual model (heterotic or otherwise)
where massless states appear at the perturbative level
so that the singularity structure can be understood at the effective field theory level.
This issue is under investigation.

Finally, one aspect that still eludes our understanding from Type I point of view
is the topological version of tadpole cancellation,
namely the condition relating the degree of the holomorphic maps and Euler character of the embedded Riemann surface.

\paragraph{Acknowledgements} We thank J.~Walcher, P.~Georgieva, E.~Witten, G.~Bonelli and E.~Gava for discussions.
In particular we thank Ashoke Sen for pointing out that
for the amplitude considered in \cite{Antoniadis:1993ze},
one should be able to choose different gauges for the bosonic and fermionic parts of the graviton vertex.
The work of NP is partially supported by the COST Action MP1210 ``The string Theory Universe''
under ECOST-STSM-MP1210-130116-068480
and by the Italian National Group of Mathematical Physics (GNFM-INdAM).
NP also thanks the IFT in Madrid and the physical mathematics group of J.~Walcher in Heidelberg
for hospitality at different stages of this work.
The work of A.T. is supported by the INFN Iniziativa Specifica GAST.

\appendix

\section{Old type IIA computation}
\label{app93}

The quantity\footnote
{In this section, we assume familiarity with the original computation\cite{Antoniadis:1993ze};
therefore we omit to carefully explain some notations and details,
and concentrate on the aspects that are new.}
we are looking for is $\mathcal F_g (\mathcal W^2)^g$.
From $(\mathcal W^2)^{g-1}$ we take the lowest components to get $(T_+ T_-)^{g-1}$.
From the remaining $\mathcal W^2$ we take $(R.T)_{\mu \nu} \theta_1 \sigma^{\mu \nu} \theta_2$
and finally take the remaining two $\theta$ from $\mathcal F_g$,
which gives $(D_a \mathcal F_g) F^a_{\mu \nu} \theta_1 \sigma^{\mu \nu} \theta_2$
where $T$ are anti self-dual graviphoton field strengths,
$R$ the anti self-dual Riemann tensor and
$F^a_{\mu \nu}$ is the anti self-dual field strength in a chiral vector multiplet $V^a$ labeled by the index $a$.\footnote
{Note that this is chiral vector multiplet and not anti-chiral.
In the latter case one would be probing holomorphic anomaly and
that would be a completely different calculation.}
Recall that $\mathcal F_g$ is a function of vector multiplets $V^a$ and so
$(D_a \mathcal F_g) = \frac{\partial \mathcal F_g(\chi)}{\partial \chi^a}$
where $\chi^a$ are the lowest components (i.e.\ moduli of Calabi-Yau) of the vector super-fields $V^a$.
Thus we have $2g-1$ graviphotons, one $R$ and one $F^a$.
All the field strength vertices are in $(-\frac12)$ picture
(we are focusing on the left moving sector --- discussion for the right moving part is identical)
so total number of PCO on genus $g$ surface is $(2g-2 + \frac{1}{2}(2g-1)+\frac{1}{2})=3g-2$.
To be explicit let us work with orbifold CY.
The internal part of the vertex for $T$ carries charge $(\frac{1}{2},\frac{1}{2},\frac{1}{2})$ in the three internal planes
and for $F^a$ we take it to be $(-\frac{1}{2},-\frac{1}{2},\frac{1}{2})$.
This means that the total internal charge of the vertices is $(g-1,g-1,g)$
therefore all the $(3g-2)$ PCOs can only contribute the internal parts of the supercurrents.
Note that SUSY transformation of $F^a$ vertex gives the vertex operator for $\chi^a$ in $(-1)$ picture with charge $(0,0,1)$,
which is the vertex of the untwisted modulus relating to
change in the complex structure (or complexified volume) in IIB (or IIA) of the third 2-torus.
The spacetime part of the spin field is as follows:
$(g-1)$ of $T$ (at points $x_i$, $i=1,\ldots,g-1$) come with $S_1$,
$g$ come with $S_2$ (at points $y_m$, $m=1,\ldots,g$)
and $F^a$ at point $w$ comes with $S_2$.
So altogether there are $(g-1)$ $S_1$ and $(g+1)$ $S_2$.
This means that the vertex for the Riemann tensor (at $z$) must be $R_{0+0+}$ i.e.\ $\psi_1 \psi_2$
to balance the space-time charges.
In other words the bosonic part of $R$ vertex does \emph{not} contribute.

With this assignment $(g-1)$ of the PCO must contribute $\bar{\psi}_3 \partial X^3$ (say at $u^{(1)}_i$, $i=1,\ldots,g-1$),
$(g-1)$ of the PCO must contribute $\bar{\psi}_4 \partial X^4$ (say at $u^{(2)}_i$, $i=1,\ldots,g-1$)
and $g$ of the PCO must contribute $\bar{\psi}_5 \partial X^5$ (say at $u^{(3)}_i$, $i=1,\ldots,g$)
where scripts $3,4,5$ on $X$ and $\psi$ refer
to the complex coordinates and their fermionic partners of the three tori respectively.

The correlation function,
apart from the prime forms (given by the OPEs) and $\sigma$'s that take care of the dimensions and monodromies,
is
\be
\frac{
\theta_s (z+\frac{1}{2}(x-y-w))^2
\,
\theta_{g_3,s}(\frac{1}{2}(x+y+w)-u^{(3)})
\,
\prod_{i=1}^2 \theta_{g_i,s} (\frac12 (x+y-w)-u^{(i)})
}
{\theta_s (\frac{1}{2}(x+y+w)-u+2 \Delta)}
\ee
where summation of $x,y,u^{(1)},u^{(2)},u^{(3)}$ is implied and
$u$ denotes the sum over all the $(3g-2)$ positions of PCO.
Now we can choose the gauge
\be
 u = y +w-z +2\Delta
\ee
then the denominator cancels with one of the spacetime $\theta$ functions.
After the spin structure sum one finds
\be
 \theta(z+x-w-\Delta)\theta_{g_1}(\Delta-u^{(1)})\theta_{g_2}(\Delta-u^{(2)})\theta_{g_3}(\Delta+w-u^{(3)})
\ee
Finally we multiply the above by (by using gauge condition)
\be
 1=\frac{\theta(y-z-\Delta)}{\theta(u-w-3\Delta)}
\ee
Now together with appropriate prime forms and $\sigma$'s
that are already there in the original correlation function
\be
\label{det}
\theta(z+x-w-\Delta) \theta(y-z-\Delta)= (\det \omega(z,x)) (\det\omega(y))
\ee
Furthermore by using bosonization and together with appropriate prime forms and $\sigma$'s
that are already there in the original correlation function
\be
 \theta_{g_1}(\Delta-u^{(1)})= \langle \prod \bar{\psi}_3(u^{(1)}) \rangle, \quad
 \theta_{g_2}(\Delta-u^{(2)})= \langle \prod \bar{\psi}_4(u^{(2)})\rangle, \\
 \theta_{g_3}(\Delta+w-u^{(3)})= \langle \psi_5(w) \prod \bar{\psi}_5(u^{(3)}) \rangle
\ee
where these are correlation functions in the twisted theory
(i.e.\ $\bar{\psi}_{3,4,5}$ are dimension one and $\psi_{3,4,5}$ are dimension zero).
Combining also $\partial X_3(u^{(1)})$, $\partial X_4(u^{(2)})$ and $\partial X_5(u^{(3)})$
and taking all partitioning of $u$ into the three groups and anti-symmetrizing,
the above becomes
\be
 \langle \psi_5(w) \prod G_-(u) \rangle
\ee
where $G_-$ is the $\NN=2$ world sheet supercurrent with $U(1)$ charge $(-1)$:
this is again in the twisted theory, i.e.\ $G_-$ has dimension 2 and $G_+$ with dimension 1 is the topological BRST current.
Note that the above correlator has first order poles as $w$ approaches any of the $u$
and has first order zeros when any of the $u$ goes to any other $u$.
Finally 
\be
 \frac{1}{\theta(u-w-3\Delta)} \rightarrow \frac{1}{\langle c(w) \prod b(u)\rangle}
\ee
where $\rightarrow$ means after taking into account various prime forms and $\sigma$'s,
$b$ and $c$ are the standard $(b,c)$ ghost system of dimension $(2,-1)$.

Now we can take one of the $u$'s (say $u_{3g-2}$) to approach $w$:
\be
\frac{ \langle \psi_5(w) \prod_{i=1}^{3g-2} G_-(u_i) \rangle}
{\langle c(w) \prod_{i=1}^{3g-2} b(u_i)\rangle}
= \partial X^5(w)
\frac{ \langle \prod_{i=1}^{3g-3} G_-(u_i) \rangle}
{\langle \prod_{i=1}^{3g-3} b(u_i) \rangle}
\ee
where we have used the OPE
\be
\psi_5(w) G_-(u_{3g-2}) =\partial X^5(w)\frac{1}{w-u_{3g-2}},
\quad c(w) b(u_{3g-2})=\frac{1}{w-u_{3g-2}}
\ee
and the fact that $\partial X^5(w)$ just gives the zero modes
(note that $G$ contains only $\partial X^5 $ and not $\partial \bar{X^5}$).
Now $\frac{ \langle \prod_{i=1}^{3g-3} G_-(u_i) \rangle}{\langle \prod_{i=1}^{3g-3} b(u_i) \rangle}$
is independent of $u_i$ as both numerator and denominator are proportional to $\det h(u)$
where $h(u)$ are the $(3g-3)$ quadratic differentials.
So
\be
\frac{ \langle \prod_{i=1}^{3g-3} G_-(u_i) \rangle}{\langle \prod_{i=1}^{3g-3} b(u_i) \rangle}
\langle \prod_{i=1}^{3g-3} (\mu_i b) \rangle
= \langle \prod_{i=1}^{3g-3} (\mu_i G_-) \rangle
\ee
where $\mu_i$ are the Beltrami differentials and $(\mu_i G_-)= \int \mu_i G_-$.
Combining also the right moving part and integrating $(z,x,y)$ using \cref{det} one finds $(\det\Im\tau)^2$,
which cancels with the contribution from the space-time $X$ zero mode integrations.
The final result is in IIB
\be
\int_{\mathcal M_g} \langle \prod_{i=1}^{3g-3} (\mu_i G_-) \prod_{i=1}^{3g-3} (\mu_i \tilde{G_-}) \int_w \partial X^5 \bar{\partial} X^5(w) \rangle
=\partial_a \mathcal F^B_g
\ee
where the derivative is w.r.t.\ complex structure moduli of the CY,
and in IIA
\be
\int_{\mathcal M_g} \langle \prod_{i=1}^{3g-3} (\mu_i G_-) \prod_{i=1}^{3g-3} (\mu_i \tilde{G_+}) \int_w \partial X^5 \bar{\partial} \bar{X}^5(w) \rangle
=\partial_a \mathcal F^A_g
\ee
where derivative is w.r.t.\ complexified K\"ahler moduli of CY.
In both the cases the derivatives is with respect to the holomorphic vector moduli as is to be expected.

All of the above can be done for an arbitrary CY (i.e.\ not necessarily orbifold)\cite{Antoniadis:1993ze}.

\section{Theta functions}
\label{theta}

Generalized $\theta$-function for genus $g$ Riemann surface $\Sigma_g$ is defined on $\IC^g$ as
\be
\theta (v|\tau)= \sum_{n \in \IZ^g} \exp \left( i\pi n \cdot \tau \cdot n + 2 i\pi n \cdot v \right)
\ee
where the positions that enter the arguments of theta functions
are defined on the Jacobi variety of $\Sigma_g$,
for example $v=\frac12 (x-y)+z$ means
$v_\mu = \int_{P_0}^z \omega_\mu + \frac12 \int_y^x \omega_\mu \in \IC^g$,
with $x,y,z \in \Sigma_g$ and $P_0$ some base point.
Here $\omega_\mu$ for $\mu=1,\ldots,g$ are the abelian differentials
and $\tau$ is the period matrix of $\Sigma_g$;
sometimes we drop $\tau$ and just write $\theta (v)$.
The generalization with spin structure $s=(a,b)\in\left(\frac12 \IZ/\IZ\right)^{2g}$ is given by
\be
\label{theta-spin-def}
\theta_s(v|\tau)=e^{i\pi a \cdot \tau a + 2 i\pi a \cdot (v+b)}\,\theta (v+\tau a+b|\tau)
\ee
while the twisted one is $\theta_{s,g}(v|\tau)=\theta_{(a,b+g)}(v|\tau)$.
Since $\theta_s(-x)=(-1)^{4 a\cdot b}\theta_s(x)$,
we distinguish accordingly between even and odd spin-structures.

Riemann vanishing theorem states that for all $z \in \IC^g$
the function $f(P)=\theta(z+ \int_{P_0}^P \omega)$
either vanishes identically for all $P \in \Sigma_g$, or it has exactly $g$ zeros $Q_i$ on $\Sigma_g$;
moreover, in the latter case, there exists a vector $\Delta \in \IC^g$, called the Riemann class, depending only on $P_0$,
such that the points $Q_i$ satisfy
$z+\sum_i \int_{P_0}^{Q_i} \omega \equiv \Delta$, modulo elements in the period lattice.
Note that $\Delta$ depends on the choice of $P_0$ in such a way that e.g.\
\be
\theta\left( \frac12 \left( \sum_{i=1}^g (x_i+\bar{x}_i) +z+\bar{z}\right) - \sum_{a=1}^{3g-3} u_a + 2 \Delta \right)
\ee
is independent of $P_0$.

A useful identity\cite[II\S6 eq.~($R_\text{ch}$) p.~214]{mumford1} due to Riemann is
\be
\label{riem-id}
2^{-g} \sum_s \theta_s(x) \, \theta_s(y) \, \theta_s(u) \, \theta_s(v)=\\
\theta \left( \frac{x+y+u+v}2 \right) \,
\theta \left( \frac{x+y-u-v}2 \right) \,
\theta \left( \frac{x-y+u-v}2 \right) \,
\theta \left( \frac{x-y-u+v}2 \right)
\ee

\bibliographystyle{JHEP}
\bibliography{biblio}
\end{document}